\pdfoutput=1
\documentclass[11pt]{article}
\usepackage[letterpaper]{geometry}
\usepackage{fullpage,graphicx,color,wrapfig}
\usepackage[tight,footnotesize]{subfigure}
\usepackage{epsfig,url,balance}
\usepackage[noadjust]{cite}
\usepackage{amsmath,amssymb,amsthm,mathtools}
\usepackage{multirow,multicol,import,fancyhdr}
\usepackage[fit,breakall]{truncate}
\usepackage{makeidx}
\usepackage{xspace}
\usepackage{float}
\usepackage[bottom]{footmisc}
\usepackage{enumitem}

\DeclareGraphicsExtensions{.pdf,.png,.jpg}

\DeclareMathOperator*{\argmax}{arg\,max}

\begin{document}

\newtheorem{definition}{Definition}
\numberwithin{definition}{section}

\newtheorem*{definition*}{Definition}

\newtheorem{theorem}{Theorem}
\numberwithin{theorem}{section}

\newtheorem{lemma}{Lemma}
\numberwithin{lemma}{section}

\newtheorem{corollary}{Corollary}
\numberwithin{corollary}{section}

\newtheorem{claim}{Claim}

\newtheorem*{notation*}{Notation}

\newtheorem{conjecture}{Conjecture}

\newcommand{\rationalOne}{minimally rational}

\newcommand{\rationalTwo}{competitively rational}

 \title{Socially Optimal Mining Pools}
\author{Ben A. Fisch\footnote{Stanford University, bfisch@stanford.edu} \and Rafael Pass\footnote{Cornell University, rafael@cs.cornell.edu} \and abhi shelat\footnote{Northeastern University, a.shelat@northeastern.edu}}
\date{}
\maketitle

\setlength{\unitlength}{5mm}

\begin{abstract}
Mining for Bitcoins is a high-risk high-reward activity. Miners, seeking to reduce their variance and earn steadier rewards, collaborate in
so-called \emph{pooling strategies} where they jointly mine for Bitcoins. Whenever some pool participant is successful, the 
earned rewards are appropriately split among all pool participants. 
Currently a dozen of different pooling strategies (i.e., methods for
distributing the rewards) are in use for Bitcoin mining.

We here propose a formal model of utility and social welfare for Bitcoin mining (and analogous mining systems) based on the
theory of discounted expected utility, and next study pooling
strategies that maximize the social welfare of miners.
Our main result shows that one of the pooling strategies actually
employed in practice---the so-called \emph{geometric pay
 pool}---achieves the optimal steady-state utility for miners when
its parameters are set appropriately.

Our results apply not only to Bitcoin mining pools, but any other form of pooled mining or 
crowdsourcing computations where the participants engage in repeated random trials towards a common goal, and where ``partial'' solutions can be
efficiently verified.
\end{abstract}

\pagebreak 

\section{Introduction}

In recent years, \emph{crowd-sourcing of computation}---where anyone can contribute to a computationally heavy task---has grown in popularity. For instance, in the  SETI@home project, users search for extraterrestrial life by analyzing radio telescoping data; or in the Rosetta@home project, users  process data to discover new proteins. In both of these examples, however, the participating users freely volunteer computing resources.
With the advent of Bitcoin, a new type of  computational crowdsourcing emerged: in place of  altruism,  users are \emph{incentivized} to participate in the computation by receiving a reward (paid in Bitcoins) for performing the work.
Bitcoin \cite{Nakamoto} is a digital currency system that enables users to transact without a central authority. In absence of a trusted central monitor, the system relies on  external monitors called ``miners'' who perform intensive  computation---searching for a solution to a computation puzzle---to operate the system. To incentivize participation, miners receive rewards for any puzzle they solve.  The reward incentive in Bitcoin has an exceedingly high variance (the puzzles are difficult to solve), and as shown in~\cite{PSS}, this is inherent in the Bitcoin system. As a result, miners typically collaborate by forming  \emph{mining pools} to reduce their variance. Currently, miners use many different types of pooling strategies. 

The focus of this work is to determine the optimal mining pooling strategies. While our focus is on Bitcoin, our results  apply to any form of mining that involves random trials and where demonstration of partial work is possible. One can even imagine applications to non-computational forms of mining (e.g., gold mining, oil drilling). The hope is that insight from analyzing the Bitcoin mining system will also be useful to incentivize other types of crowdsourced computation (such as those mentioned above). 

We begin with an overview of Bitcoin system and then proceed to formalize the pool-design problem. We refer the reader to~\cite{SoK:Bitcoin} for a more detailed description of the Bitcoin system.

\subsection{Overview of Bitcoin} 
\paragraph{The Bitcoin reward system} Bitcoin uses a distributed consensus protocol to maintain in a public ledger called the \textit{blockchain} which stores  the valid transaction history. Participants broadcast transactions over a peer-to-peer network, while agents called \textit{miners} collect blocks of transactions, verify their integrity, and append them to the blockchain. Miners are required to produce a computationally intensive \textit{proof-of-work} in order to append a  block to the blockchain. This mechanism makes it difficult for malicious miners controlling less than a majority of the system's computational power to rewrite history (as long as the mining hardness is appropriately set \cite{GKL, PSS}). The system incentivizes miners by rewarding them with newly minted coins for each block they add to the chain.

The proof-of-work consists of finding a partial pre-image for a cryptographic hash function $H$. Roughly, given a block with contents $b$, a miner must find a value $r$ from a large domain $X$ such that $H(b || r) < d$. Miners successively sample random values in $X$ until they find a solution to this cryptographic puzzle. The value $d$ determines the block \textit{difficulty}, or the probability $p$ that a random $r \in X$ will satisfy the puzzle. The current difficulty is set so that in expectation, the entire group of miners succeed in mining a single block every 10 minutes (and as shown by the analysis in \cite{PSS}, the mining difficulty cannot be significantly decreased without making the protocol vulnerable to attacks.)

As a consequence, based on the hardness parameters for 2015, an individual miner with state-of-the-art mining hardware will in expectation mine a single block once every 687 days~\cite{Lewenberg}! Moreover, the process of mining is memoryless. A miner who has not received any reward after 687 days must still wait another 687 days on average to receive a reward. Thus, the income of an individual miner has a very high variance. The number of blocks produced by a miner working at a continuous rate $h$ (measured in hashes per second) for a time period $t$ is well approximated by a Poisson distribution with mean $\lambda = p h t$. The miner receives expected reward $\lambda B$ with variance $\lambda B^2$, where $B$ is the reward per block. 

\paragraph{Mining pools.} Miners seeking to reduce their variance and earn steadier incomes join \textit{mining pools}. Participating in a pool is called \textit{pool mining} and mining alone is called \textit{solo mining}. Whenever a pool miner wins a reward, the reward is shared among all the pool's participating miners.  Pools require a trusted operator to monitor participation and manage the allocation of rewards. 
The rough idea is to have the pool operator monitor how much work each individual participant contributes to the pool, and then whenever some participant manages to mine a block, the operator receives the block reward in proxy and then allocates the reward among the pool participants based on how much work they contributed. 

Monitoring the effort of participating miners, however, is a nontrivial task. Unless miners are assumed to be honest,
simply asking miners to report their effort leads to \emph{free riding}: riders will claim to have done work even if they have not.
To overcome this problem, miners instead demonstrate their effort by submitting partial proofs-of-work called \textit{shares}, which are simply block hashes that satisfy a lower difficulty parameter, i.e. shares are a ``near-solution'' to the original computational puzzle. We distinguish such shares from full solutions which we simply refer to as \emph{blocks}. 

To prevent pool participants from stealing the block-mining reward whenever they find a full solution, the block owner identity is incorporated into the proof-of-work. Pools only accept proofs-of-work, partial or complete, that incorporate the identity of the pool as the block owner. Otherwise, miners could submit only partial proofs to the pool and send their complete proofs to the Bitcoin network for a solo reward.

The principal question we consider now is:
\begin{quote}
\emph{How should a pool operator allocate block rewards to the pool participants so as to maximize their social welfare?}
\end{quote}
If miners are \emph{risk neutral}, then solo mining is  optimal. But if miners are \emph{risk averse} (technically, have a concave utility function), then pooling strategies may improve their utility by decreasing the variance of their rewards.
From here on, we refer to the pool's strategy for allocating the reward as the \emph{allocation rule}. 
Indeed, several popular pooling strategies with different allocation rules are currently in use:\footnote{See  \url{https://en.bitcoin.it/wiki/Comparison_of_mining_pools} for a
  comprehensive list of popular pools.}

\begin{itemize} 
\item In the \emph{proportional pay} scheme, the reward of a block is 
split among all the participants in the pool 
proportionally to the number of shares they submitted to the pool---in other words, the rewards are split evenly among the shares in the pool, and the pool is then ``emptied'' for the next round. 
\item The \textit{Pay-Per-Last-N-Shares} (PPLNS) pool is similar, except that the block reward is always distributed evenly among the last $N$ shares submitted to the pool (without ever ``emptying'' the pool). 

\item Score based pooling mechanisms generalize PPLNS pools, and distribute block rewards over preceding shares contributed to the pool according to some weighting function. PPLNS can be viewed as a score based pooling mechanism that uses a step weighting function. Rewards in the \textit{Slush's pool} and the \textit{geometric pool} are concentrated at the winning block and decay exponentially over the preceding shares.
\end{itemize} 


Some pools do not allocate rewards immediately, and instead invest rewards in a central pool fund. These funds may be used to incentivize future participation in the pool at a risk-free rate (\textit{Pay-Per-Share} (PPS) pools). PPS pools absorb all the variance of their participants, and in order to survive with high probability they must heavily discount the risk-free rate. This is not a pure pooling mechanism because it assumes a financier. In this paper, we restrict our attention to pure pooling mechanisms.

\begin{definition} [Informal Definition]  A \emph{pure pooling mechanism} is an allocation rule that assigns fractional rewards to all shares preceding a block, including the block share itself. The allocation-rule may depend on the state of the pool. 
\end{definition}

\subsection{Our Results}
In its current state of affairs, the Bitcoin mining pool ecosystem is a collection of seemingly ad-hoc mining pool strategies with ad-hoc parameters, and there is no consensus as to which pool mechanism is ``optimal''. As far as we are aware, there are no published theories on optimal mining pool strategies even among a restricted class of strategies. Towards this goal, we put forward a formal model of utility of pooling strategies for computation/mining, and derive the pooling strategy that maximizes miners' utility.
We demonstrate that for the most commonly used utility function, a power utility function, the \emph{geometric pool} is optimal if the parameters of the geometric pool are appropriately set. As mentioned above, the geometric pool is one that is used in practice (although not necessarily with the optimal parameters).




\paragraph{Modeling mining pools.} In order to analyze the question of what the optimal pooling strategy is, we must first specify a model for measuring the utility of a pool participant. We start by viewing Bitcoin mining pool shares as financial investments that receive a cashflow from the pool. Different pools represent different investment packages, each varying the risk, value, and timing of payoffs. According to the \textit{expected utility} (EU) theory, risk-averse agents have a concave utility function, most commonly the power utility $u(x) = x^\alpha$, and prefer assets $X$ that maximize the expected utility $E[u(X)]$. We use the standard \textit{discounted expected utility} (DEU) \cite{Samuelson} model for ranking the utility of sequences of time-separated payoffs, which exponentially discounts utilities occurring $t$ steps in the future by $\delta^t$ for some constant discount parameter $\delta$.

DEU is well suited for modeling the utility of a single share that a miner contributes to a pool. However, when an individual miner contributes multiple shares to one or more pools over time, the utility of the shares is not in general separable (see \S\ref{sub:utility_pool_share} for a discussion).  However, we argue that the utility of shares is approximately separable when the miner controls a sufficiently small portion of the pool's computational power. It is reasonable to restrict our attention to large pools of relatively small miners. The core principle of mining pools is that small miners with high mining reward variance join large pools in order to reduce their reward variance.




\paragraph{Defining Optimality.} 
Our goal is to find pooling strategies that maximize the \emph{social welfare} of all miners.  How do we measure social welfare?
A general pool may, for example, distribute rewards in an arbitrary way that benefits some miners at the expense of others. 
In contrast, in a \textit{perfectly fair} pooling strategy, all shares receive \emph{equal} utility. 
Many natural pool strategies don't quite achieve perfect fairness, but do achieve \emph{steady-state fairness}, 
where the expected utility of pool shares converges to a \emph{steady-state utility} in the lifetime of the pool. 
For example, in a PPLNS pool the first N shares earn a higher expected reward than all shares, but all subsequent shares have the same ``steady-state" expected reward/utility. 
Likewise, geometric pools are steady-state fair, although convergence does not occur in finitely many steps as in PPLNS. 

Since our goal is to focus on the social welfare of the pool miners as a whole, we exclude pools that do not at least achieve fairness in the limit, i.e. pools that do not have a well defined steady state utility. We define the \emph{social-welfare} of a steady-state pool to be its steady-state utility. We refer to a pool as \emph{steady-state optimal} if it achieves the \emph{optimal} steady-state utility. 

\begin{definition} [Informal Definition]  The \emph{social welfare} of a pool is the pool's steady-state utility if it exists, and 0 otherwise. 
\end{definition}

\paragraph{Main Theorems.}
In general, pools may have a complex reward allocation that depends on the pool's state including the history of prior reward allocations in the pool. In practice, all the pooling strategies that are used in practice except for proportional pay  are significantly simpler: they use a \emph{fixed rule} that is independent of the history of the pool to allocate rewards to the miners who contributed previous shares to the pool. We can represent such fixed-rule pools by an infinite length vector $\vec{X}$ where $X_i$ denotes the fractional reward allocated to the miner who contributed the $i$th share preceding a reward-earning block. 

\begin{definition}
A \textbf{fixed-rule pool} is a pool that has a fixed allocation rule such that whenever a block reward is earned the pool distributes a fixed fraction $X_i$ of the reward to the $i$th pool share preceding the block share, where $i \geq 0$. 
\end{definition}

Fixed-rule pools are indeed preferable, and as we show in our first theorem, we can limit our study to such objects without any loss of generality.   In particular,  for \emph{any} concave utility function, if there exists a steady-state optimal \emph{fixed-rule} pool (i.e., a specific fixed-rule pool that is optimal among the set of fixed-rule pools), then this pool is also steady-state optimal among all pools.

{
\renewcommand{\thetheorem}{\ref{thm:steady_to_fixed}}
\begin{theorem}
For any concave real-valued utility function $u$, time-discounting parameter $\delta < 1$, and block reward $B$, if there exists a steady-state optimal fixed-rule pool then this pool is steady-state optimal. 
\end{theorem}
\addtocounter{theorem}{-1}
}

\begin{figure}[ht!]
\centering\includegraphics{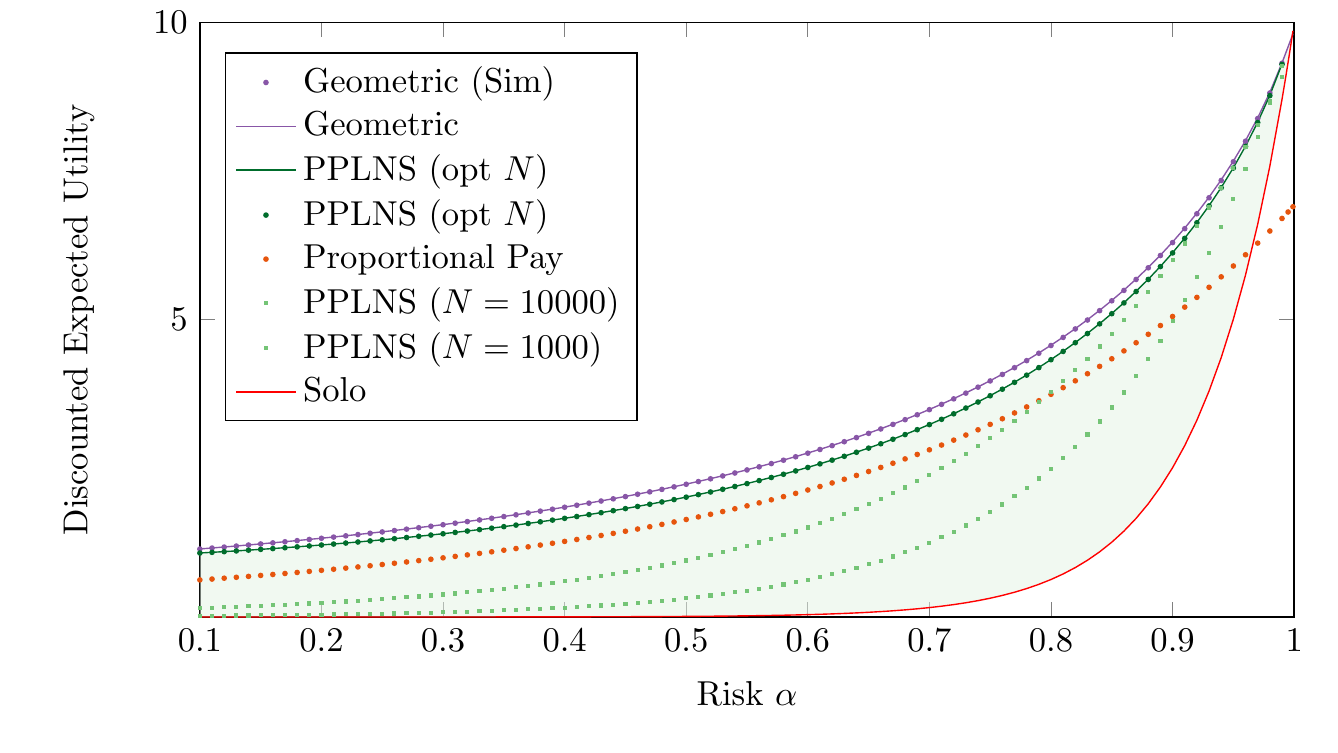}
 \caption{Expected value of each share for various mining pool schemes as a function of risk tolerance $\alpha$ for the power utility function $u(x)=x^\alpha$.  The win rate is $p=10^{-5}$, discount rate $d=0.99999$, and reward $B=10^6$.  Dotted lines represent simulated data, smooth lines represent analytically-derived results.  The area in green represents the range for PPLNS ranging from $N=1$ (solo) to the optimal values for $N$ for a given $\alpha$.
}\label{fig-all-pools}
\end{figure}

In our main theorem, we then characterize an optimal  steady-state pooling strategy for  a common family of utility functions.  In Fig.~\ref{fig-all-pools}, we illustrate Thm.~\ref{thm:optimal_power_utility} by graphing the results of simulating each type of pool for 1 billion shares and then computing the discounted expected utility for each share as a function of the miner's utility function, i.e., the miner's risk parameters $\alpha$.  Each experiment was run 50 times, and the dots reflect the average of each experiment, whereas the solid lines represent our analytical results.   

{
\renewcommand{\thetheorem}{\ref{thm:optimal_power_utility}}
\begin{theorem}
For the utility function $u(x) = x^\alpha$ where $0 < \alpha < 1$, the geometric pool with allocation rule $X_i = B(1 - \delta^{1/1-\alpha}) \delta^{i/1-\alpha}$ is steady-state optimal. 
\end{theorem}
\addtocounter{theorem}{-1}
}

\begin{figure}[ht!]
\centering\includegraphics{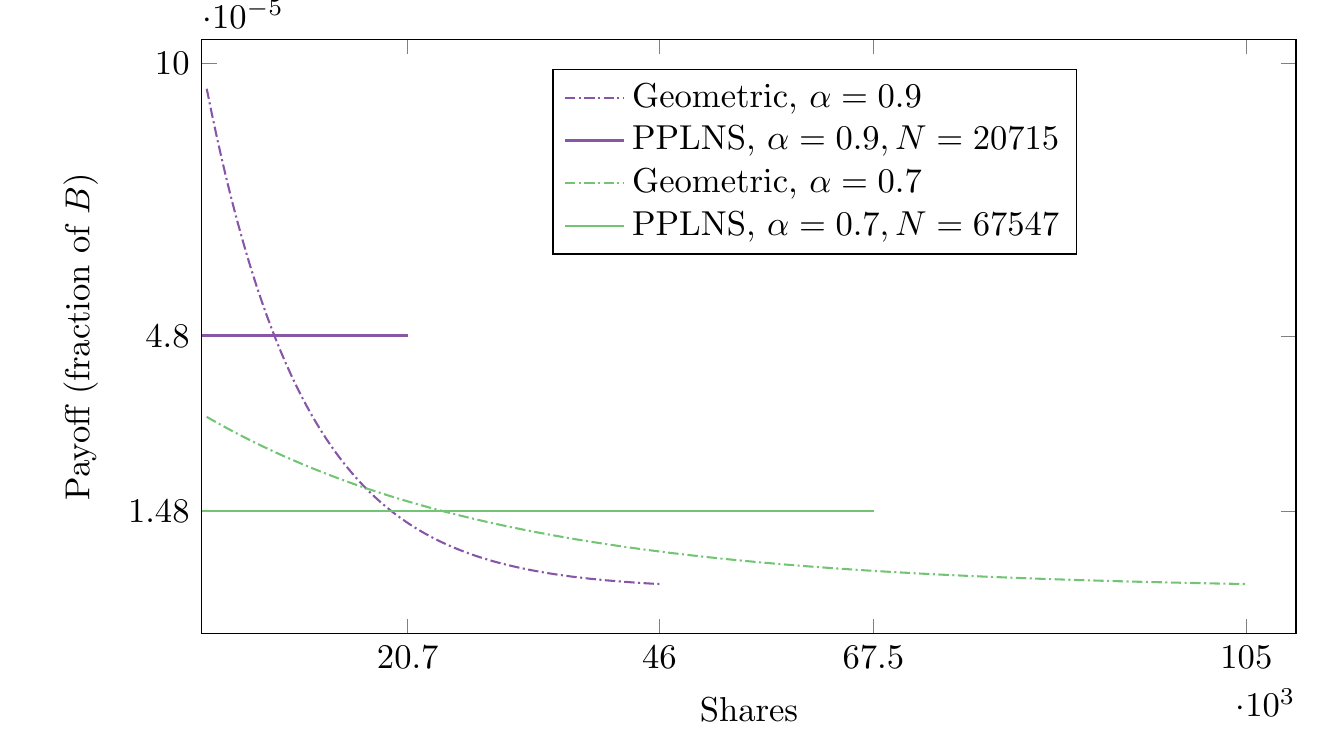}
\caption{Share payoff fraction for geometric scheme and PPLNS.  The geometric scheme pays
some shares more than PPLNS, but pays many more shares less than PPLNS. 
The PPLNS schemes depicted use optimal choices of $N$ for the risk parameters and for the power utility function; the win rate is $p=10^{-5}$, discount rate $d=0.99999$.
}
\label{fig-geom-vs-pplns}
\end{figure}

Intuitively, the geometric pools achieves a slightly higher DEU per share than the optimal pay-per-last-$N$ scheme because it distributes the block reward in a non-uniform manner, with some shares receiving more rewards than PPLNS, and many many more shares receiving less.  Fig.~\ref{fig-geom-vs-pplns} shows sample payoff structures for a few parameters of these schemes.

\paragraph{Incentive compatibility} The question of incentive compatibility in mining pools has been addressed to some degree in several other works \cite{Rosenfeld, Lewenberg, Schrijvers, Laszka, Eyal15}, but in general remains largely open. The well-known counterexample to incentive compatibility is the proportional pay pool, which is vulnerable to \emph{pool hopping}. The optimal pool we derive is not vulnerable to pool hopping \cite{Rosenfeld}. Block rewards in a proportional pay pool are distributed evenly among all shares submitted per block period. Thus, as the block period length (number of shares in a block period) increases, the utility of contributing shares to the pool decreases. If a block period grows beyond a certain point, participating miners achieve a better expected utility by solo mining (or mining for a different pool). There are several other known attacks in the context of competing pools. Some of these attacks enable miners to boost their rewards at the expense of other more honest participants \cite{Eyal15, Laszka, Johnson}.

While current research suggests that achieving incentive compatibility in a system of competing pools is extremely difficult, one can even ask the simpler question of whether a single pool is incentive compatible (i.e., with respect to deviations to solo-mining). In fact, the optimal pool that we derive is incentive compatible in this sense. The steady-state utility of the optimal pool is, by virtue of being optimal, superior to the utility of solo mining. Furthermore, since the pool share utility in the optimal pool is monotonic non-increasing in its convergence to the steady-state utility, the pool share utility is always better than the utility of solo mining. Hence, miners participating in this pool will never be incentivized to switch to solo mining at any point in time.

\section{The Utility Model}\label{sec:mining_utility}
\subsection{Mining Pools}

A miner invests work in repeated attempts to solve a computationally difficult puzzle in order to win a prize. After every repeated attempt, the miner learns whether or not the attempt was successful. Previous attempts do not affect future attempts, and thus, at every renewed attempt the miner has the same probability of receiving an award. This is similar to a player in a scratchcard lottery who repeatedly purchases cards, scratching off each card before purchasing the next. If every card purchase is a Bernoulli trial with success parameter $p$, then the number of wins out of $N$ trials has a Binomial distribution with expectation $pN$ and variance $p(1-p)N$. 

\paragraph{Monitoring mining work.} In Bitcoin mining, the analog of a scratchcard purchase is an investment of work. Just as a scratchcard lottery pool operator would count purchased cards, Bitcoin mining pool operators monitor the \textit{work} of their participating miners. 
Currently, operators estimate participants' work rates by collecting partial proofs-of-work called shares. Producing a share is significantly easier than producing a block, but sufficiently difficult so that miners cannot feasibly produce shares without honestly attempting to produce a valid block. 

\paragraph{Rewarding shares.} The pool operator collects shares in an inherently sequential manner, and we assume that the history of shares submitted to the pool is common knowledge among all participants. Each share wins a reward with independent probability $p$. 

A \textit{pool mechanism} is a rule for distributing block rewards over past and future shares. The reward of an individual share is a sum over rewards it receives from past or future shares, as well as any reward it generates and keeps for itself when it is a valid block. 

Formally, we define a reward allocation rule as a probabilistic function of the pool's state. The pool's state includes the history of shares contributed to the pool and their outcomes (i.e. partial or valid block). We can denote this state $\sigma = (t, h_t)$, where $t$ is the number of shares, and $h_t$ is a binary vector of length $t$ indicating if each previous share was a block. The output of the allocation rule is a collection of random variables denoting reward payments to specific shares (i.e. the miners who contributed those shares). We restrict our definition to pure pooling strategies in which the reward is immediately allocated to miners who previously contributed shares to the pool (i.e no future payments). 

\begin{definition}\label{def:allocation_rule}
An \textbf{allocation rule} is a function $\mathcal{A}(t, h_t) = \{X^{(t, h_t)}_i\}_{0 \leq i \leq t}$ where $X^{(\sigma)}_i$ is a random variable denoting the value allocated to the contributor of the pool's $i$th share when the pool wins a block reward $B$ in state $\sigma = (t, h_t)$. 
\end{definition}

\subsection{Risk-Aversity and Time-Discounting}

\paragraph{Risk-averse utility.} Expected utility (EU) theory gives a way to order preferences of consumption balancing expected value and risk. Given a random variable $X$ representing the value of a consumption, EU calculates assigns $X$ an ordinal value by calculating the expectation $E[u(X)]$ for a \textit{von Neumann-Morgenstern utility} function $u$. The function $u$ maps consumptions to the real line, and determines the marginal increase in utility as the quantity of a consumption increases. A von Neumann-Morgenstern utility function only uniquely characterizes second order behavior; it is only defined up to affine linear transformations. One common family of utility functions is the power utility $u(x) = x^\alpha$ for $\alpha > 0$. Each function in this family is increasing on $(0, \infty)$ and $\alpha$ determines its concavity/convexity. There is a strong connection between $\alpha$ and the risk-aversity of an agent. A risk-neutral agent will have a linear utility function. When $u$ is linear, $E[u(X)] = u(E[X])$, and thus the ordinal value of $X$ is only dependent on its expected value. A risk-averse agent will have a concave utility function. When $u$ is concave, $E[u(X)] \leq u(E[X])$ by Jensen's inequality and $E[u(X)]$ monotonically decreases as the variance of $X$ increases for fixed $E[X]$.

\paragraph{Time-discounting.} When sharing their risk across time-separated shares, miners need a way to assess the present value of their shares. Time-discounting captures factors such as the waiting time for future cash flow and the risk of pool termination. Consider as an example the PPLNS strategy. Miners commit to paying the preceding $N-1$ shares a $1/N$ fraction of their block reward $B$ if they win. The miner expects only $pB/N$ in immediate compensation for his work, but assuming the miners producing the next $N$ shares do the same, the miner's expected reward is $pB$. The variance of each payment is $p(1-p)(B/N)^2$, and since the payments are independent the total variance is $p(1-p)B^2/N$, which is a factor $1/N$ lower than the solo mining variance. It would seem that as $N$ becomes infinitely large the variance disappears while the expected reward remains the same. However, the miner also expects to wait an infinite amount of time to recover any noticeable reward. 

\paragraph{Discounted expected utility.} The classic model for calculating the expected total utility of intertemporal consumptions is the \textit{discounted expected utility} (DEU) model. The DEU formula calculates a present expected utility of time-separated consumptions $(c_1,...,c_k)$, where each $c_t$ occurs in the time period $t$, and $u$ is a von Neumann-Morgenstern utility function:

$$DEU(c_1,...,c_k) = \sum_{t = 0}^k E[u(c_t)] \delta^t$$

The value $\delta < 1$ is at time-discounting parameter, and is generally chosen to be close to 1. Note that the DEU summation converges even as $k \rightarrow \infty$. There are many implicit axioms in the DEU model formula, see \cite{FrederickLoewenstein} for a comprehensive overview. In particular, since the DEU model treats utility as linearly additive over time-separated consumptions, it implicitly assumes that the consumer is risk-neutral to aggregated utilities over time, even if the consumer is risk-averse in each time period. 
Intertemporal risk-aversion has also been considered in the economics literature and there are modified DEU models where aggregation of discounted utilities is nonlinear \cite{FrederickLoewenstein, EpsteinZin, KrepsPorteus}.

\subsection{Player Utility}\label{sub:utility_pool_share}

\paragraph{Validity of DEU.} For the DEU model to be valid, the time separation between events must be sufficiently high. Cashflows in Bitcoin mining pools come from shares, but the time separation between individual pool shares may be quite small depending on the size of the pool. For simplicity we will assume that the time separation between pool shares, which is determined by the difficulty parameter of mining a share and the number of users in the pool, is sufficiently high. More generally, to address the issue of time-separation between shares we can bundle pool shares into periods. The events of the DEU sum would be cashflows occurring in successive periods. 

The DEU model is well suited for modeling the utility of a
single share that a miner contributes to a pool, but how do we account for multiple shares submitted by the same miner? If the utilities of a miner's separate shares are additively separable then the miner's overall utility is simply the sum of the individual utilities. However, in general, a miner's pool shares are not separable. As an extreme example, consider a miner who controls the entire pool and simulates the proportional pay pool over its shares, where each share earns an equal fraction of the total reward. Looking only at individual shares gives the false illusion that the miner now achieves a higher utility. In reality, the miner has decreased the variance of each individual share's earnings by increasing the covariance of all the share earnings. 

Thus, in order for the DEU approach to be valid for multiple shares submitted by the same miner, we additionally assume that each miner does a \emph{small} amount of work compared with the rest of the pool. Specifically, we assume that the rate at which any individual miner contributes shares to the pool is small compared to the overall pool share rate. This is to ensure that the (time discounted) cashflow to a pool share from other shares submitted by the same miner is negligible compared to the (time discounted) cashflow received from shares submitted by the other miners. 
This is essentially \textit{multiselves approach}, which assumes that miners act as a new independent agent each time they contribute a new share to the pool, seeking to maximize utility in that moment.

\paragraph{DEU of a pool share.} Let the random variable $X_i$ denote the reward that the share accrues during the $i$th period following the submission of the share. The variable $X_0$ is the reward that the share generates and keeps for itself.

\begin{definition}\label{def:utility}
Given discount parameter $\delta$ and utility function $u$, the discounted expected utility of a pool share is:

$$U =  \sum_{i \geq 0} E[u(X_i)] \delta^i$$

\end{definition}

\subsection{Pool Optimality} 

We say that a pool strategy achieves \textit{steady-state fairness} if the utility of contributing pool shares converges with overwhelming probability. 

\begin{definition}\label{def:steady_state_fair}
A pool strategy is \textbf{steady-state fair} if the sequence $\{U_k\}$ converges in $\mathbb{R}$, where $U_k$ denotes the expected utility of the $k$th pool share. The limit point of $\lim_{k \rightarrow \infty} U_k$ is the \textbf{steady-state utility} of the pool. 
\end{definition}





\paragraph{Steady-state optimality.} 

In any class $\mathcal{C}$ of steady-state fair strategies, we can define the steady-state optimal strategies of $\mathcal{C}$ as the set of strategies in $\mathcal{C}$ that have the highest steady-state utility. 

\begin{definition}
A pool strategy $\mathfrak{p}$ is \textbf{steady-state optimal} for a class $\mathcal{C}$ of steady-state fair pool strategies if and only if
$ \mathfrak{p} \in \argmax_{x \in \mathcal{C}} \lim_{k \rightarrow \infty} E[U(x)_k]$.
\end{definition}

\section{The Optimal Pool}\label{sec:optimal_pool}

In this section we show how to derive a steady-state optimal pool for honest risk-averse players. The parameters of the optimal pool will depend on the choice of utility function $u$, time-discounting factor $\delta$, and fixed block reward $B$. The main results of this section applies to any general utility function $u$ that is concave and real-valued. We first show a relationship between steady-state optimal pools and \emph{fixed-rule pools}--a pool that allocates block rewards according to a fixed rule, independent of the pool's state. Specifically, we prove that if there exists an optimal fixed-rule pool then it is also steady-state optimal. The optimal fixed-rule pool is a solution to a convex optimization problem that depends on $u$, $\delta$, and $B$. We solve this optimization problem explicitly for the power utility function $u(x) = x^\alpha$ ($0 < \alpha  < 1$), which yields a geometric pool whose parameters depend on $\alpha$. 

Our results are the following three theorems: 

\begin{theorem}\label{thm:steady_to_fixed}
For any concave real-valued utility function $u$, time-discounting parameter $\delta < 1$, and block reward $B$, if there exists an optimal fixed-rule pool then this pool is steady-state optimal. 
\end{theorem} 
\begin{proof} 
In Lemma~\ref{lem:steady_state_bounded_fixed_pool} we show that if a pool has steady-state utility $U$ then for every $\epsilon$ there exists a fixed-rule pool that has utility at least $U - \epsilon$. Therefore, $U$ is bounded by the supremum of fixed-rule pool utilities. If there exists an optimal fixed-rule pool then by definition it achieves this supremum and hence its utility is an upper bound on the steady-state utility of any steady-state fair pool. 
\end{proof}

\begin{theorem}\label{thm:optimal_fixed_general_form}
There exists an optimal fixed-rule pool if and only if there is a solution to the following convex optimization problem:

\begin{align*}
\argmax_{\vec{x}}{\sum_{i \geq 0} u(x_i) \delta^i} \ \ \textnormal{subject to} \ \
\sum_{i \geq 0} x_i \leq B, \forall i \ x_i \geq 0
\end{align*}

If it exists, the solution $\{x_i\}_{i \geq 0}$ is the optimal fixed-rule pool.
\end{theorem}

\begin{theorem}\label{thm:optimal_power_utility}
For the power utility functions there is a fixed-rule geometric pool that is steady-state optimal. The parameters of this geometric pool are determined by the block reward $B$, the risk-aversity parameter $0 < \alpha < 1$ of the utility function $u(x) = x^\alpha$, and the time-discounting factor $\delta$. Specifically, this geometric pool has the allocation rule $X_i = B(1 - \delta^{1/1-\alpha}) \delta^{i/1-\alpha}$. 
\end{theorem}

\subsection{Fixed-rule pools}

Fixed-rule pools have several nice properties: they are perfectly fair, and the expected reward of any share in the pool is bounded. 
  
\begin{claim}\label{claim:fixed_rule_bound}
For fixed allocation rules $\sum_{t \geq k} E[X^{(t)}_k] \leq B$ for any $k$. 
\end{claim}
\begin{proof} 
Suppose towards contradiction that $\sum_{i \geq 0} E[X^{(k+i)}_k] = \hat{B} > B$ for some $k$. By definition of a limit, for any $\epsilon > 0$ there exists $N_\epsilon$ such that $|\sum_{i = 0}^{N_\epsilon} E[X^{(k+i)}_k] - \hat{B}| < \epsilon$. Setting $\epsilon = (\hat{B} - B)/2$ implies $\sum_{i = 0}^{N_\epsilon} E[X^{(k+i)}_k] > B$. However, using the property of fixed allocation rules, $\sum_{i = 0}^{N_\epsilon} E[X^{(k+i)}_k] = \sum_{i = 0}^{N_\epsilon} E[X^{N_\epsilon}_i] \leq B$. This is a contradiction.

\end{proof}

In a pool with a fixed allocation rule $\{X_i\}$, we can express the utility of any share in terms of the allocation rule variables as follows: 

\begin{equation}\label{eqn:fixed_utility}
U = \sum_{i \geq 0} pE[u(X_i)] \delta^i
\end{equation}

Since every share has the same expected utility, by definition the pool is perfectly fair. 

\begin{claim}\label{claim:fixed_rule_fairness}
Every pool with a fixed allocation rule is perfectly fair. 
\end{claim}

\subsubsection{Proof of Theorem~\ref{thm:optimal_fixed_general_form}}

The utility of any share in a fixed-rule pool with allocation rule $\{X_i\}$ is $\sum_{i \geq 0} pE[u(X_i)] \delta^i$ for variables $X_i \geq 0$ where $\sum_{i \geq 0} E[X_i] \leq B$ (Claim~\ref{claim:fixed_rule_bound}). Therefore, the optimal fixed-rule pool is the solution to:
\begin{align*}
\argmax_{\vec{X}}{\sum_{i \geq 0} E[u(X_i)] \delta^i} \ \ \textnormal{subject to} \ \
\sum_{i \geq 0} E[X_i] \leq B, \forall i \ E[X_i] \geq 0
\end{align*}

By Jensen's inequality, for concave $u$ we have $E[u(X_i)] \leq u(E[X_i])$, and equality holds when $X_i$ are scalars or $u$ is linear. 
Thus it suffices to solve the optimization for scalars $x_i$ as follows:

\begin{align*}
\argmax_{\vec{y}}{\sum_{i \geq 0} u(y_i) \delta^i} \ \ \textnormal{subject to} \ \
\sum_{i \geq 0} y_i \leq B, \forall i \ y_i \geq 0
\end{align*}

If a solution exists then it defines the allocation rule of an optimal fixed-rule pool. Conversely, if some pool with allocation rule $\{X^*_i\}$ is optimal, then the pool with allocation rule $\{E[X^*_i]\}$ is necessarily optimal, hence it is a solution to the above optimization problem. 

Finally, to show that this is a convex optimization problem we will prove that the objective function $f(\vec{y}) = \sum_i u(y_i) \delta^i$ is concave. Since $u$ is concave, for any $\vec{y}^{(1)}$, $\vec{y_2}^{(2)}$
 and scalar $t$ it holds that 
 $f(t \vec{y}^{(1)} + (1-t)\vec{y}^{(2)}) 
 = \sum_i u(t y^{(1)}_i + (1-t)y^{(2)}_i) \delta^i 
 \leq \sum_i t u(y^{(1)}_i)\delta^i + (1-t) u(y^{(2)}_i)\delta^i 
 = t f(\vec{y}^{(1)}) + (1-t) f(\vec{y}^{(2)})$.

\subsection{Steady-state pools to fixed-rule pools}

\begin{lemma}\label{lem:steady_state_bounded_fixed_pool}
For any steady-state fair pool $\mathfrak{p}$ that has steady-state share utility $U_\mathfrak{p}$ and any $\epsilon > 0$ there exists a fixed-rule pool $\mathfrak{p}'$ 
that has share utility $U_{\mathfrak{p}'} \geq U_\mathfrak{p} - \epsilon$. 
\end{lemma}

\subsubsection{Proof of Lemma~\ref{lem:steady_state_bounded_fixed_pool}}

In fixed-rule pools the distribution of future rewards a miner receives for submitting a share was independent of state. In more general pools, even steady-state fair pools, this distribution of future rewards could fluctuate over the state of the pool. The high level idea of this proof is to show that if the utility of the pool converges then the distribution of future expected rewards converges in some subsequence of states to a fixed distribution. We use this fixed distribution of expected rewards to define a fixed-rule pool that allocates to each previous share exactly its expected reward. The steady-state utility of this subsequence of states will be bounded by the utility of this fixed-rule pool. Since infinite subsequences of any convergent sequence also converge to the same limit, it follows that the steady-state utility of the pool is also bounded by the utility of this fixed-rule pool. 

If the space containing the sequence of expected reward distributions were sequentially compact, then existence of a subsequence of states for which reward distributions converge would follow immediately. However, each expected reward distribution is an infinite vector over $\mathbb{R}$, and infinite dimensional subspaces of $\mathbb{R}^\infty$ are not necessarily sequentially compact. Instead, we examine \emph{finite-window} pools, which only allocate rewards over preceding shares within some finite window. Due to time-discounting, every pool can be approximated by a finite-window pool. More precisely, for any pool we can define the finite-window pool that uses the same allocation rule restricted to the last $N$ shares, and for any $\epsilon > 0$ we can choose $N$ sufficiently large so that the utility of any share in the finite-window pool is within $\epsilon$ of the same share in the original pool. In finite-windows pools the expected reward distribution is a finite length vector in a closed and bounded subset of $\mathbb{R}^N$, which by the Bolzano-Weierstrass theorem is sequentially compact. Thus, we can prove that the utility of all finite-window pool approximations are bounded by the utility of a fixed-rule pool, and by making $\epsilon$ arbitrarily small we extend this bound to the original pool. 

\begin{definition} 
A pool has \textbf{finite window n} if its allocation rule $\mathcal{A}(t, h_t) = \{X^{(t, h_t)}_i\}_{0 \leq i \leq t}$ satisfies $X^{(t, h_t)}_i = 0$ for all $i \leq t - n$. A \textbf{finite-window pool} has a finite window n for some $n \in \mathbb{N}$. 
\end{definition} 

\begin{notation*} $ $
\begin{itemize}
\item $U^{(\sigma)}$ denotes the expected utility of a share submitted to the pool in state $\sigma = (t, h_t)$. 
\item $U^t$ denotes the a priori expected utility of the $t$th share submitted to the pool, i.e. $U^t = E[U^{(t, h_t)}]$ over the distribution of $h_t$. 
\item $\vec{r}^t = (r^t_0,...,r^t_{d - 1})$ denotes an expected reward vector of the $t$th share in a pool of finite window $d$ where $r^t_i$ denotes that the a priori expected reward the $t$th share receives from the $t + i$th share. 
\item $R^t$ denotes the a priori expected reward of the $t${th} share submitted to the pool. In a pool of finite window $R^t = |\vec{r}^t|_1 = \sum_i r^t_i$. 
\item Given a pool $\mathfrak{p}$ let $\mathfrak{p}^{(d)}$ denote the finite-window pool that truncates the allocation rule of $\mathfrak{p}$ to the last $d$ shares, i.e. in any state $\mathfrak{p}^{(d)}$ has the same allocation $X^{(t, h_t)}_i$ as $\mathfrak{p}$ for $t - d < i \leq t$ and $X^{(t, h_t)}_i= 0$ for all $i \leq t - d$. 
\end{itemize}
\end{notation*}

\begin{definition} 
Let $\mathfrak{p}$ and $\mathfrak{p}'$ be two pools with $k$th share utility $U_k$ and $U'_k$ respectively. The pool $\mathfrak{p}'$ \textbf{$\epsilon$-approximates} the pool $\mathfrak{p}$ if $|U_k  - U'_k| \leq \epsilon$ for all $k \geq 1$. 
\end{definition}

For the following claims we assume the utility function $u$ is concave and real-valued. 
\begin{claim}\label{claim:finite_window_epsilon_approximation}
For any pool $\mathfrak{p}$ and $\epsilon > 0$, there exists $d$ such that the finite window pool $\mathfrak{p}^{(d)}$ $\epsilon$-approximates $\mathfrak{p}$. 
\end{claim} 
\begin{proof} 
If in some state $\sigma$ the pool $\mathfrak{p}$ has utility $U^{(\sigma)} = \sum_{i \geq 0} E[u(X^\sigma_i)]\delta^i$ then in the same state $\mathfrak{p}^{(d)}$ has utility $U'^{(\sigma)}_d = \sum_{i = 0}^{d-1} E[u(X^\sigma_i)] \delta^i$. The random variable $X^\sigma_i$ is the reward the share submitted in state $\sigma$ receives from the $i$th succeeding share, so $E[X^\sigma_i] \in [0, B]$. Furthermore, since $u$ is concave $E[u(X^\sigma_i)] \leq u(E[X^\sigma_i])$ by Jensen's inequality and $u$ is continuous and bounded on $[0, B]$ by some value $\beta$. We can therefore apply the same argument used in Claim~\ref{claim:uniform_convergence_truncated_functions} to show that for any $\epsilon$ there exists $d$ such that $|U^{(\sigma)} - U'^{(\sigma)}_d|  =  |\sum_{i = d}^ \infty E[u(X^\sigma_i)] \delta^i | \leq \sum_{i = d}^\infty \beta \delta^i < \epsilon$. 

\end{proof}

\begin{claim}\label{claim:finite_window_pools_are_balanced} 
In finite-window pools the limit of means of expected reward is bounded by $pB$, i.e. $\lim_{n \rightarrow \infty} (1/n) \sum_{t =1}^n  R^t \leq pB$. 
\end{claim}
\begin{proof} 
Consider any pool with some finite window $d$. A share in this pool can receive rewards only the $d$ shares that immediately succeed it. Therefore, all rewards allocated to the first $n$ shares in the pool come from the first $n + d - 1$ shares. Each share contributes at most $B$ to this total reward, and only if it is a valid block. In expectation only a $p$ fraction of these blocks will be blocks. Therefore, the expected total reward received by the first $n$ shares is at most $p(n + d - 1) B$.

This implies $\lim_{n \rightarrow \infty} (1/n) \sum_{t =1}^n  R^t \leq \lim_{n \rightarrow \infty} (1/n) p(n+d - 1)B = p B$. 
\end{proof}

\begin{claim}\label{claim:balanced_subsequence}
Any pool in which the limit of means of expected reward is bounded by some value $\mathcal{B}$ there exists a subsequence of shares in which the expected reward of every share is bounded by $\mathcal{B}$.
\end{claim}
\begin{proof} 
Suppose towards contradiction that this claim is false. 
Then there exists a $k$ such that for all $k' \geq k$ the reward of the $k'$th share is greater than $\mathcal{B}$. This implies
$\lim_{n \rightarrow \infty} (1/n) \sum_{t = 0}^n R^t 
\geq \lim_{n \rightarrow \infty} (1/n) \sum_{i = k +1}^n R^t 
> \lim_{n \rightarrow \infty} (1/n) (n - k + 1) \mathcal{B}  = \mathcal{B}$, 
which contradicts the hypothesis that the limit of means of expected reward is bounded by $\mathcal{B}$. 
\end{proof}

\begin{claim}\label{claim:reward_convergence_finite_window_pools}
In any finite-window pool there exists a subsequence $\{t_k\}$ of shares such that the expected reward vectors $\vec{r}^{t_k}$ converge to a vector $\vec{r}^*$ such that $|\vec{r}^*|_1 \leq pB$.
\end{claim}
\begin{proof} 
Consider a pool of finite window $d$. By Claim~\ref{claim:finite_window_pools_are_balanced} and Claim~\ref{claim:balanced_subsequence} we can find a subsequence $\{t_i\}$ of shares whose expected rewards are all bounded by $pB$. Thus the sequence of reward vectors $\{\vec{r}^{t_i}\}$ all lie in the closed ball $\{x : |x|_1 \leq pB\} \subseteq [0, pB]^{d+1}$. This space is compact, hence by the Bolzano-Weierstrass theorem $\{\vec{r}^{t_i}\}$ has a convergent subsequence in this space. 
\end{proof}



\begin{claim}\label{claim:finite_window_fixed_pool_bound}
The steady-state utility of any steady-state finite-window pool is bounded by the utility of some fixed-rule pool. 
\end{claim}
\begin{proof}
Consider any steady-state pool of finite window $d$. Let $\vec{r}^* \in \mathbb{R}^{d}$ be the limit point of the convergent subsequence $\{t_k\}$ of expected reward vectors in the pool guaranteed by Claim~\ref{claim:reward_convergence_finite_window_pools}. 
Define the fixed-rule pool with allocation rule $X_i = r^*_i / p$ for $0 \leq i \leq d - 1$ and $X_i = 0$ elsewhere. This is a valid allocation rule because $\sum X_i = (1/p) |\vec{r}^*|_1 \leq B$. We will now show that the utility of this fixed pool, $U_{\textsf{fixed}}$ is an upper bound on the steady-state utility $U^*$ of the finite-window pool. 

The utility of the shares in any infinite subsequence of a steady-state pool must converge to the pool's steady-state utility (infinite subsequences of any convergent sequence in $\mathbb{R}$ converge to the same limit). Thus we have both $\lim_{k \rightarrow \infty} U^{t_k} = U^*$ and $\lim_{k \rightarrow \infty} \vec{r}^{t_k} = \vec{r}^*$. For any $\epsilon$ we can choose $K$ sufficiently large so that both $|U^{t_K} - U^*| < \epsilon$ and $|r^{t_K}_i - r^*_i| < \epsilon$ for all $i$. 

Let $Y_i$ denote the reward that the $t_K$\textsuperscript{th} share receives from the $(t_K + i)$\textsuperscript{th} share conditioned on the event that this share is a block. 
By Jensen's inequality we can bound $U^{t_K}$ as 
\begin{align*}
U^{t_K} &= \sum_{i = 0}^{d - 1} p\cdot E[u(Y_i)] \cdot\delta^i 
\leq \sum_{i = 0}^{d-1} p \cdot u(E[Y_i]) \cdot \delta^i  
 = \sum_{i=0}^{d-1} p \cdot u(r^{t_K}_i / p) \\
&\leq \sum_{i= 0}^{d-1} p \ u((r^*_i + \epsilon) / p)
\end{align*}

Since $U_{\textsf{fixed}} = \sum_{i=0}^{d-1} p \ u(r^*_i / p) \delta^i$ we have:
$$U_{\textsf{fixed}} - U^* \geq U_{\textsf{fixed}} - U^{t_K} -  \epsilon 
\geq \left(\sum_{i = 0}^{d-1} p (u(r^*_i / p) - u((r^*_i + \epsilon) / p)) \delta^i \right) - \epsilon$$

Since $u$ is continuous, $\lim_{\epsilon \rightarrow 0} |u((r^*_i + \epsilon) / p) - u(r^*_i / p)| = 0$. 
Therefore, for any $\epsilon' > 0$ we can show that $U^* \leq U_{\textsf{fixed}} + \epsilon'$, hence $U_{\textsf{fixed}}  \leq U^*$. 

\end{proof}

From Claim~\ref{claim:finite_window_epsilon_approximation}, for any steady-state pool $\mathfrak{p}$ and for any $\epsilon$ there exists a finite-window pool $\mathfrak{p}'$ that $\epsilon$-approximates $\mathfrak{p}$. By Claim~\ref{claim:finite_window_fixed_pool_bound} the steady-state utility of $\mathfrak{p}'$ is bounded by the utility $U_{\textsf{fixed}}$ of a fixed-rule pool. Therefore, the steady-state utility $U^*$ of $\mathfrak{p}$ satisfies $U^* \leq U_{\textsf{fixed}} + \epsilon$. This concludes the proof of Lemma~\ref{lem:steady_state_bounded_fixed_pool}.

\subsection{Optimal pool for power utility}


\subsubsection{Proof of Theorem~\ref{thm:optimal_power_utility}}
 
The main challenge in this proof is that the optimization problem is over $\mathbb{R}_{\geq 0}^\infty$ rather than $\mathbb{R}_{\geq 0}^n$. Constrained convex optimization over infinite dimensional spaces is in general nontrivial. First let us define the following notation: 

\begin{notation*} 

Define $f(y) = \sum_{i \geq 0} u(y_i) \delta^i$ and $g(y) = \sum_{i \geq 0} y_i$  for $y \in \mathbb{R}_{\geq 0}^\infty$ and $n \in \mathbb{N}$, 
where $u$ is the concave utility function in question.
For $n \geq 1$ define the ``truncated" sums $f_n(y) = \sum_{i=1}^n u(y_i) \delta^i$ and $g_n(y) = \sum_{i=1}^n y_i$. The functions $f_n(y)$ and $g_n(y)$ 
are well defined over both $\mathbb{R}_{\geq 0}^\infty$ and $\mathbb{R}_{\geq 0}^n$. 

\end{notation*}

Our approach is to solve for a maximizer $x^*_n$ of each $f_n(y)$ subject to $g(y) \leq B$ and $\forall i \ y_i \geq 0$ (Claim~\ref{claim:optimal_strategy_truncated_fixed}). To obtain some $x^*_n$, it suffices to solve for a maximizer of $f_n(y)$ defined instead over $y \in \mathbb{R}_{\geq 0}^n$ with the constraint $g_n(y)$, and then extend this maximizer to a point in $\mathbb{R}^\infty$ that is identical to this maximizer in the first $n$ components and 0 in every other component (Claim~\ref{claim:extend_truncated_maximizer_constraint}). The solutions $x^*_n$ are obtained via the method of Lagrange multipliers. We then show that this sequence of maximizers converges in $\mathbb{R}^\infty$\footnotemark, i.e. $\{x^*_n\} \rightarrow x^*$, and the limit point $x^*$ is a maximizer of $f(y)$ subject to $g(y) \leq B$ over $\mathbb{R}^\infty$ (Claim~\ref{claim:maximizer_convergence}, Claim~\ref{claim:uniform_convergence_truncated_functions}). 

\footnotetext{Convergence in $\mathbb{R}^\infty$ can be defined with respect to the standard Euclidean norm restricted to points in $\mathbb{R}^\infty$ that have finite norm. All the points in the sequence of maximizers lie in this subspace because they have a finite number of nonzero components. The limit point of this sequence satisfies the optimization constraint (i.e. has a bounded L1 norm) and thus also lies in this subspace.}

\begin{claim}\label{claim:optimal_strategy_truncated_fixed} When $u(x) = x^\alpha$, the maximizer of $f_n(y)$ subject to $g_n(y) \leq B$ over $\mathbb{R}_{\geq 0}^n$ is $y_i = B\frac{1 - \delta^{1/1-\alpha}}{1- \delta^{n/1-\alpha}} \delta^{i/1-\alpha}$.  \end{claim}

\begin{proof}
$f_n(y)$ is increasing in $y_i$ for every $i$. Thus, if there exists a global maximum then it is achieved on $g_n(y) = B$. The Lagrangian for this optimization is $L(y, \lambda) = f_n(y) + \lambda(B - g_n(y))$. There exists a solution $y^*$ to the constrained optimization problem if and only if there exists $\lambda^*$ such that $L(y^*, \lambda^*)$ is a global maximum of the Lagrangian. \textbf{First, we prove that a solution exists} by examining the principal minors of the Hessian of the Lagrangian. A solution exists if for all $k \geq 2$, the determinant of the $k$th principal minor of the Hessian of $L(y, \lambda)$ has sign $(-1)^{k + 1}$.  \textbf{Second, we derive a unique stationary point of the Lagrangian}. A global maximum $(y^*, \lambda^*)$ must be a stationary point of the Lagrangian, i.e. $\nabla_y L(y^*,\lambda^*) = 0$.  Since the stationary point we derive is unique it must be the global maximum of the Lagrangian, and hence a solution to the constrained optimization. 

\paragraph{Existence of a solution.} The Hessian of $L(y, \lambda)$ is:

\begin{equation*}
\begin{pmatrix}
    0  & \frac{\partial{g_n}}{\partial x_1} & \cdots  & \frac{\partial{g_n}}{\partial x_n} \\
    \frac{\partial{g_n}}{\partial x_1}  & \ddots & & \\
    \vdots & & \frac{\partial^2{L}}{\partial x_i \partial x_j} & \\
     \frac{\partial{g_n}}{\partial x_n} & & & \ddots
\end{pmatrix} = 
\begin{pmatrix}
    0  & 1 & \cdots  & 1 \\
   1 & -\ell_{1,1} & & \\
    \vdots & & \ddots & \\
    1 & & & -\ell_{n,n}
\end{pmatrix}
\end{equation*}

 For any $i \neq j$ we have $\frac{\partial{L}}{\partial{y_i} \partial {y_j}} = 0$ and for any $i$ we have $\frac{\partial^2{L}}{\partial{y_i^2}} = \delta^i \frac{\partial^2{u(y_i)}}{\partial{y_i^2}} < 0$ by the strict concavity of $u(y_i)$. Additionally, $\frac{\partial{g_n}}{\partial x_i} = 1$ for all $i$. Consider the $k$th principal minor $M^{(k)}$. 
The Leibniz formula for the determinant of $M^{(k)}$ is $\det (M^{(k)}) = \sum_{\sigma \in S_k} sgn(\sigma) \prod_{i = 1}^k m_{i, \sigma_i}$. 
Consider any nonzero term of this sum. It cannot include in its product $m_{1, 1}$ and so must contain $m_{1, i} = 1$ and $m_{j, 1} = 1$ for some $i, j \neq 1$. The $k - 2$ remaining elements of the product must be from the nonzero (negative valued) diagonal entries. However, it also cannot include the elements $m_{j, j}$ and $m_{i, i}$ because it includes $m_{1, i}$ and $m_{j, 1}$. If $i \neq j$ this leaves only $k - 3$ diagonal elements, hence necessarily $i = j$ and the product includes all the $k - 2$ diagonal elements except $m_{i, i}$. The term corresponds to an odd permutation $\sigma$ that contains $k - 2$ fixed points and a single inversion $(1, i)$, so $sgn(\sigma) = -1$. 
All nonzero terms thus have sign $-(-1)^{k-2} = (-1)^{k -1} = (-1)^{k +1}$. Therefore $\det (M^{(k)})$ has sign $(-1)^{k +1}$.

\paragraph{Unique stationary point.} We proceed to show that the Lagrangian $L(y, \lambda) = f_n(y) + \lambda(B - g_n(y))$ has a unique stationary point when $u(x) = x^\alpha$. We will first do this for $u(x) = x^\alpha$. Setting $\nabla_y L(y,\lambda) = 0$, this yields the system of equations $\alpha y_i^{\alpha-1} \delta^i - \lambda = 0 \ \textnormal{for all} \ i$, and solving for $y_i$:
\begin{align*}
y_i = \left(\frac{\alpha}{\lambda} \delta^i\right)^{1/1-\alpha}
\end{align*}
Applying the constraint $\sum_{i=1^n} y_i = B$ we get: 
\begin{align*}
\sum_{i=0}^n (\frac{\alpha}{\lambda}\delta^i)^{1/1-\alpha} 
= \left(\frac{\alpha}{\lambda}\right)^{1/1-\alpha} \sum_{i=0}^n \delta^{i/1-\alpha}
= \left(\frac{\alpha}{\lambda}\right)^{1/1-\alpha} \frac{1 - \delta^{n/1-\alpha}}{1 - \delta^{1/1-\alpha}} = B
\end{align*}
Solving for $\alpha/\lambda$ and plugging into $y_i$: 
\begin{align*}
\alpha/\lambda = \left(B\frac{1 - \delta^{1/1-\alpha}}{1- \delta^{n/1-\alpha}}\right)^{1-\alpha} 
\ \ , \ \  y_i = B\frac{1 - \delta^{1/1-\alpha}}{1- \delta^{n/1-\alpha}} \delta^{i/1-\alpha}
\end{align*}

\end{proof}

\begin{claim}\label{claim:extend_truncated_maximizer_constraint}
Let $x^* \in \mathbb{R}^n$ be a global maximizer of $f_n(x)$ subject to $g_n(x) \leq B$ over $\mathbb{R}_{\geq 0}^n$. Define $\bar{x}^* \in \mathbb{R}_{\geq 0}^\infty$ so that $\bar{x}^*_i = x^*_i$ for all $0\leq i \leq n$ and $\bar{x}^*_i = 0$ for all $i > n$. Then $\bar{x}^*$ is a global maximizer of $f_n(x)$ subject to $g(x) \leq B$ over $\mathbb{R}_{\geq 0}^\infty$. 
\end{claim}

\begin{proof} 
Consider any $x \in \mathbb{R}_{\geq 0}^\infty$ such that $g(x) \leq B$. Let $x' \in \mathbb{R}^n$ be identical to $x$ in the first $n$ components so that $f_n(x) = f_n (x')$. The constraint $g(x) \leq B$ implies that $g_n(x') \leq B$ since $x$ does not have any negative components. Hence $f_n(x) = f_n(x') \leq f_n(x^*) = f_n(\bar{x}^*)$. 
\end{proof}


\begin{claim}\label{claim:maximizer_convergence}
Let $\{f_n\}$ be a sequence of continuous real-valued functions on a metric space $X$ and $f_n \rightarrow f$ uniformly on a subset $E = \{x: g(x) = B\}$. If $\{x^*_n\}$ is a sequence such that each $f(x^*_n)$ is the global maximum of $f_n$ on $E$ and $x^*_n \rightarrow x^*$, then $f(x^*)$ is a global maximum of $f$ on $E$.  
\end{claim} 
\begin{proof}
By uniform convergence, for any $\epsilon$ there exists $n_\epsilon$ such that $\forall_y \forall_{n \geq n_\epsilon} |f_{n}(y) - f| < \epsilon$. 
Furthermore, the limit of a uniformly convergent sequence of functions is continuous, hence $f$ is continuous. Together with convergence of $x^*_n \rightarrow x^*$ this implies that for any $\epsilon$ there exists $n'_\epsilon$ such that $\forall_{n \geq n'_\epsilon} |f(x^*_{n})  - f(x^*)| < \epsilon$. Let $N = max(n_\epsilon, n'_\epsilon)$. 

Consider any $x \neq x^*$ such that $g(x) = B$. It holds that $f(x) \leq f_N(x) + \epsilon \leq f_N(x^*_N) + \epsilon \leq f(x^*_N) + 3\epsilon \leq f(x^*) + 4\epsilon$.  
Since $\epsilon$ is arbitrary, it follows that $f(x) \leq f(x^*)$. We conclude that $f(x^*)$ is the global maximum of $f$ on $E$. 

\end{proof}

\begin{claim}\label{claim:uniform_convergence_truncated_functions}
For any concave $u$, the functions $f_n(y) = \sum_{i=1}^n u(y_i) \delta^i$ converge uniformly to $f(y) = \sum_{i =1 }^\infty u(y_i) \delta^i$ on the set $E = \{ y: 0 \leq y_i  \leq B\}$. 
\end{claim}
\begin{proof} 
Since $u$ is concave it is continuous on $[0, B]$ and therefore bounded on $[0, B]$ by compactness. Let $\beta$ be the bound for $u$ on $[0, B]$. For any $y \in E$ it holds that $y_i \in [0, B]$, hence $u(y_i) \leq \beta$. 
At any point $y$, we thus have the bound
 $|\sum_{i \geq 0} u(y_i)\delta^i - \sum_{i = 1}^n u(y_i)\delta^i| = |\sum_{i \geq n} u(y_i) \delta^i| \leq u(B) \sum_{i \geq n} \delta^i = u(B) \frac{\delta^n}{1- \delta}$. 
For any $\epsilon$ there exists $N_\epsilon$ such that $\delta^n < \frac{(1-\delta)}{u(B)} \epsilon$ for all $n \geq N_\epsilon$. 
Hence, $|f_n(y) - f(y)| < \epsilon$ for all $n \geq N_\epsilon$ and all $y \in E$. 

\end{proof}

\begin{claim}\label{claim:utility_continuity}
For any concave $u: \mathbb{R} \rightarrow \mathbb{R}$, the function $f_n(x) = \sum_{i=1}^n u(x_i) \delta^i$ is continuous on $\mathbb{R}^\infty$ for any $n \in \mathbb{N}$. 
\end{claim}
\begin{proof} 
Define the function $u^{(i)} : \mathbb{R}^\infty \rightarrow \mathbb{R}$ as $u^{(i)}(x) = u(x_i)$. For any $x, x' \in \mathbb{R}^\infty$ where $|x - x'| < \delta$ it must hold that $|x_i - x'_i| < \delta$ for all $i$. Thus, since $u$ is continuous for any $\epsilon$ there exists $\delta$ such that $|x - x'| < \delta$ implies $|u^{(i)}(x) - u^{(i)}(x')| = |u(x_i) - u(x'_i)| < \epsilon$ for all $i$. 

The function $f_n(x) = \sum_{i = 1}^n u^{(i)}(x)$ is a finite sum of continuous functions, hence it is continuous. 
\end{proof}

Claim~\ref{claim:uniform_convergence_truncated_functions} and Claim~\ref{claim:utility_continuity} together show that $f_n \rightarrow f$ uniformly on the set of points in $\mathbb{R}^\infty_{\geq 0}$ that satisfy the constraint $g$. We will now show that the sequence of maximizers of $f_n$ derived in Claim~\ref{claim:optimal_strategy_truncated_fixed} converge to a point $x^*$,  and thus by Claim~\ref{claim:maximizer_convergence} $f(x^*)$ is the global maximum of $f$ on $\mathbb{R}^\infty_{\geq 0}$ subject to $g$. 

For power utility $u(x) = x^\alpha$:
\begin{align*}
x^*_i = \lim_{n \rightarrow \infty} B\frac{1 - \delta^{1/1-\alpha}}{1- \delta^{n/1-\alpha}} \delta^{i/1-\alpha} = B(1 - \delta^{1/1-\alpha}) \delta^{i/1-\alpha}
\end{align*}

\section{Future Work}\label{sec:future}
\subsection{Incentive compatibility} 

The question of incentive compatibility in mining pools has been addressed to some degree in several other works \cite{Rosenfeld, Lewenberg, Schrijvers}, but in general remains largely open. An important question relevant to our work is whether the \emph{optimal} pool we derive is incentive compatible in an environment of competing pools. 

While current research suggests that achieving incentive compatibility in a system of competing pools is extremely difficult, one can even ask the simpler question of whether pools are incentive compatible with respect to solo-mining deviations. We sketch here one possible approach. Given our assumption of small miners, we could analyze the mining pool game using the \textit{multiselves approach}, where each player is viewed as a collection of independent selves and acts as a new agent in every move of the game.  A pool $\mathfrak{p}$ is \textit{truthful} if the strategy profile in which every player mines in the pool in every state of the game is a \textit{subgame perfect Nash equilibrium} (SPE) of any game where $\mathfrak{p} \in \mathcal{P}$. We can define payoffs to participating miners in terms of our utility model for pool shares. We could show in this framework that the proportional pay pool is not an SPE by the standard technique of backwards induction. Roughly, if there exists some (apocalyptic) state over random coin tosses for which the utility of pool mining drops below the utility of solo mining, then miners in any state that immediately precedes a likely apocalypse would preemptively drop out of the pool; this logic can be recursively traced back to the first miner in the pool. 

Furthermore, it is worth mentioning that in this simple pool vs solo mining repeated game, the geometric optimal pool we derived in this work would be an SPE strategy. Note that the share utility in a geometric pool is monotonically decreasing in its convergence to steady-state. Since solo mining is a valid steady-state pool and is considered in the optimization, by definition the optimal steady-state utility will be greater than the solo mining utility. Thus, in every step of the iterated game the utility of participation in the steady-state optimal geometric pool is greater than the utility of solo mining, so there will be no deviations from this strategy. 

\subsection{Pools with investment strategies} 

Thus far we have restricted our attention to \textit{pure} pools that immediately allocate all earned rewards. Suppose that we allowed pools to maintain a budget and allocate rewards to future shares. For example, consider a \emph{Pay-Per-Next-N-Shares} (PPNNS) pool, the reverse of a $PPLNS$, where a block reward $B$ is distributed equally over the $N$ shares following a block. At the time a share is submitted to the pool, it receives a risk-free payment from previous shares that depends on the outcomes of the previous shares. The expected utility of a share is the a priori expected utility of this risk-free payment. With respect to optimizing steady-state utility, pools should only pay future shares since the utility of backward payments to previous shares are time-discounted. In fact, as $N \rightarrow \infty$, the steady-state utility in a $PPLNS$ pool approaches the utility $u(pB)$, which is optimal for balanced pools. This is because the probability a share far in the future obtains its expected reward converges to 1. 

However, as $N \rightarrow \infty$, the time for the pool to reach steady-state also grows to infinity. Early miners earn close to nothing. Although we have not yet introduced a formal model of incentive compatibility, it is intuitive that this pool would never start without some outside investment to kickstart the pool. Absent an external financier, we could imagine introducing a minimal backward payment so that the expected utility of mining any share for the pool even in times when the pool has not accumulated any rewards is at least better than solo mining. The remainder of the rewards will be allocated to future shares via forward payments. Yet, even in this hybrid forward/backward payments pool, there is no optimal strategy. We can always spread future payments thinner to increase the steady-state utility. Thus, while our model for maximizing utility was appropriate for pure pooling strategies, clearly it must be amended in order to deal with pools that make investments. 

\bibliographystyle{alpha}
\bibliography{bib/bibliography} 

\appendix
\section{Pay-Per-Last-N-Shares}\label{optimal_pplns}
 
In a PPLNS pool with parameter $N$, the block reward is allocated evenly over the last $N$ shares submitted to the pool including the valid block share, each share receiving reward $B/N$.\footnote{In practice, various techniques are used to adjust the value of $N$ to compensate for changes in the block difficulty, but to the best of our knowledge there is no consideration for choosing the optimal $N$ given $\alpha$ and $\delta$.} When $N =1$, this strategy is equivalent to solo mining. Intuitively, as $N$ increases the variance of an individual share's reward decreases while its maturity time increases. 

\paragraph{Steady-state utility.} We examine the steady-state utility of a share in a PPLNS pool:
 
 \begin{equation}\label{eq:pplns_utility}
 \sum_{i  = 1}^N pB^\alpha \left(\frac{1}{N}\right)^\alpha \delta^i = pB^\alpha \left(\frac{1}{N}\right)^\alpha \frac{1 - \delta^N}{1 - \delta}
 \end{equation}
 
 
\begin{figure}[h]\label{fig:pplns_utility}
\centering
\includegraphics[scale=0.50]{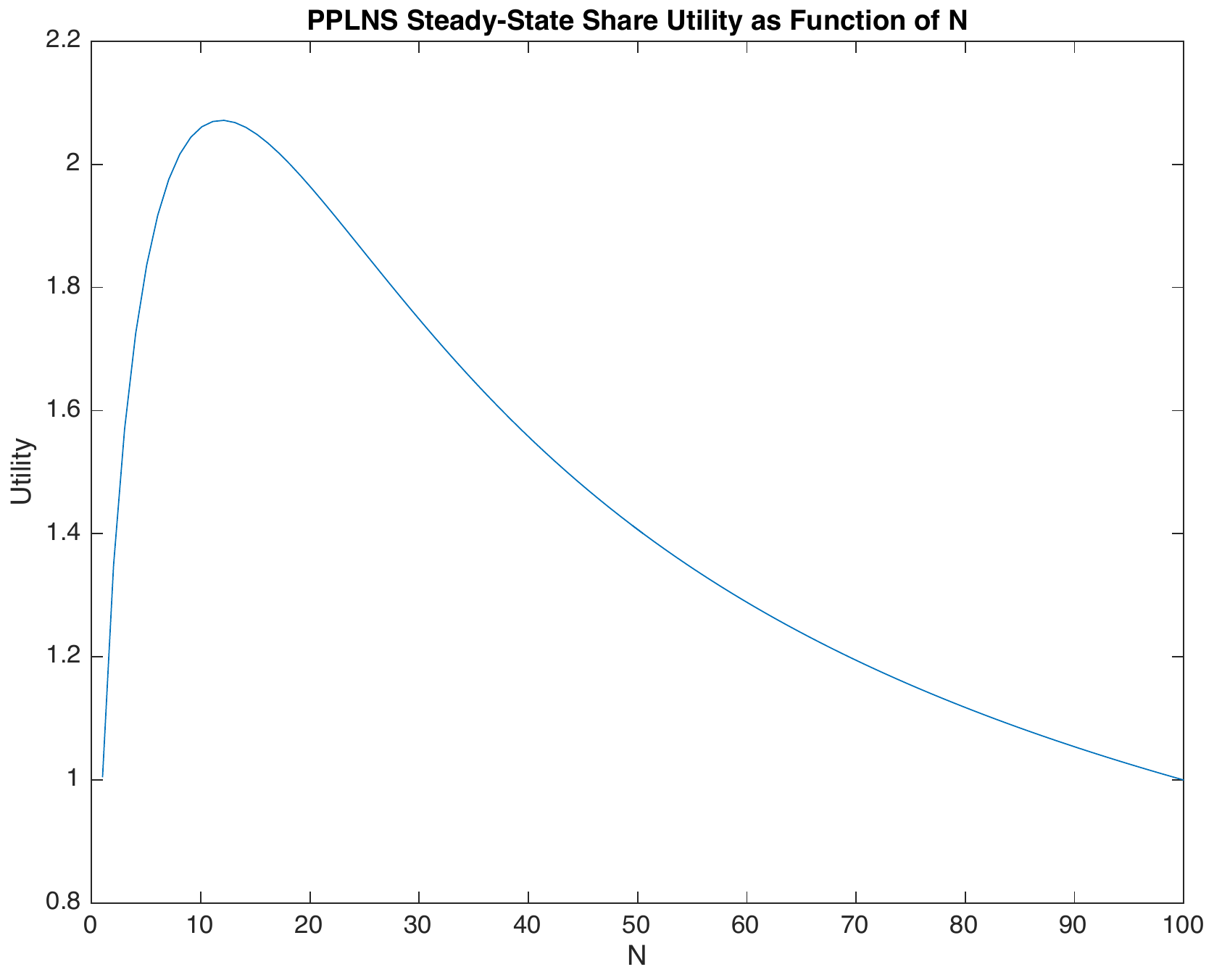}
\caption{The PPLNS utility from Equation~\ref{eq:pplns_utility} as a function of $N$ for fixed $\delta = 0.99$.}
\end{figure}

 \begin{theorem}\label{thm:optimal_PPLNS}
 Given risk-aversity parameter $\alpha < 1$ and discount parameter $\delta < 1$, the steady-state optimal PPLNS pool sets the value of $N$:
 $$N = \frac{1}{\log \delta}(W_{-1}(-e^{-\alpha} \alpha) + \alpha)$$ 
 where $W_{-1}$ is the lower branch of the product log (Lambert W) function, i.e. real valued solutions $y$ to $ye^y = xe^x$ for $y \leq -1$ and $x \in (-1, 0)$. 
 When $\alpha = 1$, the optimal PPLNS sets $N = 1$, which is equivalent to solo mining.
 \end{theorem}
 \begin{proof}
 The optimal value of $N$ maximizes the function $f(N) = (1/N)^\alpha(1-\delta^N)$. We start by examining the first derivative $f'(N)$:
 \begin{equation*}
 \begin{split}
 \frac{d f(N)}{dN} & = \frac{1}{1-\delta} (-\alpha N^{-\alpha - 1} (1 - \delta^N) - N^{-\alpha} \delta^N \log \delta) \\
 & = \frac{1}{(1- \delta)N^{\alpha +1}} (-\alpha(1- \delta^N) - N \delta^N \log \delta)
 \end{split}
 \end{equation*}
 On the interval $(0, \infty)$, we see that $f'$ is positive near $N = 0$ since: 
 \begin{equation*}
 \begin{split}
 \lim_{N\rightarrow 0} \frac{1}{(1- \delta)N^{\alpha +1}} (-\alpha(1- \delta^N) - N \delta^N \log \delta) \\
= \lim_{N \rightarrow 0} \frac{-\delta^N \log \delta}{(1- \delta) N^\alpha} 
 = \lim_{N \rightarrow 0} \frac{\log \delta^{-1}}{(1 - \delta) N^\alpha} = + \infty
 \end{split}
 \end{equation*}
 
 Additionally, we see that $f'$ is both decreasing and approaching $0$ as $N \rightarrow \infty$:
 \begin{equation*}
 \lim_{N \rightarrow \infty} \frac{1}{(1- \delta)N^{\alpha +1}} (-\alpha(1- \delta^N) - N \delta^N \log \delta) \\
= \lim_{N \rightarrow \infty} \frac{-\alpha}{(1- \delta) N^{\alpha+1}} = 0 
 \end{equation*} 
 
 Therefore, $f$ achieves at least one local maximum in the interval $(0, \infty)$. Since $\frac{1}{(1- \delta)N^{\alpha +1}} \neq 0$, the points where $f' = 0$ must satisfy:
 \begin{equation*}
 \begin{split}
 N \delta^N \log \delta = -\alpha(1 - \delta^N) 
 \Leftrightarrow N = \frac{\alpha}{\log \delta} (1 - \delta^{-N}) 
 \Leftrightarrow \delta^{-N} + \frac{\log \delta}{\alpha} N - 1 = 0 
 \end{split}
 \end{equation*}

 Define $y = \log \delta N - \alpha$, so $y$ is a linear function of $N$ and takes values in $(-\infty, -\alpha)$. Substituting $N = \frac{1}{\log \delta}(y + \alpha)$ into the above relation gives: 
 \begin{equation*}
 \begin{split}
 & \delta^{-\frac{1}{\log \delta}(y + \alpha)} + \frac{\log \delta}{\alpha} \frac{1}{\log \delta}(y + \alpha) - 1 = 0 \\
 &  \Leftrightarrow e^{-(y + \alpha)} + \frac{y + \alpha}{\alpha} - 1 = 0
 \Leftrightarrow e^{-(y + \alpha)} = -\frac{y}{\alpha} \\
 & \Leftrightarrow y e^y = -\alpha e^{-\alpha} 
 \end{split}
 \end{equation*}
 Since $-\alpha e^{-\alpha} \in [-1/e, 0]$ we can write $y = W(-\alpha e^{-\alpha})$ where $W$ is the multivalued inverse function of $x \mapsto x e^x$ defined on $[-1/e, \infty)$. $W$ is injective on $[0, \infty)$, $W(-1/e) = -1$, and $W$ and has exactly two values at every point in $(-1/e, 0)$, the principal value where $W \geq -1$ and the lower branch value where $W < -1$. The lower branch of $W$ is denoted $W_{-1}$. The principal value of $W(-\alpha e^{-\alpha})$ is $-\alpha$. Thus, the unique solution to $y$ in the interval $(-\infty, -\alpha)$ is $W_{-1}(-\alpha e^{-\alpha})$, and we obtain the unique solution for $N$: 
 
$$ N = \frac{1}{\log \delta}(W_{-1}(-\alpha e^{-\alpha}) + \alpha) $$

Since this value of $N$ is the unique zero of $f'$ in $(0, \infty)$ it must be a global maximum on this interval.

The preceding analysis was only valid for $\alpha < 1$. When $\alpha = 1$, the quantity $(1/N)^\alpha(1-\delta^N)$ is strictly decreasing on $(1, \infty)$, hence the maximum is at $N =1$. 

\end{proof}

\paragraph{Dependency of N on block difficulty.} In current practice, some PPLNS pools choose $N$ to depend on the block difficulty, setting $N \approx 1/p$. The optimal N predicted by our model is independent of p, the difficulty parameter. This might seem strange, because p determines the variance, and with higher variance it seems that N should be larger. 

The independence from p is due to the DEU's time separability assumption (in this case that utility of sequential shares that an individual miner contributes are separable). This modeling assumption implies that the improvement in utility that miners get from splitting rewards over multiple shares is independent of the individual variance of each share. This assumption is invalid if the time between shares is small or p is close to 1, but for small miners it is reasonable to assume that p is very small and the time between shares is fairly large. So, in our model, the value of N is determined only by the values of $\alpha$ (risk aversity, which encourages increasing N) and $\delta$ (time discounting, which encourages lowering N).

\section{Definitions}
 
\begin{definition*}[Social welfare]  The \textbf{social welfare} of a pool is the pool's steady-state utility if it exists, and 0 otherwise. 
\end{definition*}

\begin{definition*}[Fixed-rule pool]
A \textbf{fixed-rule pool} is a pool that has a fixed allocation rule such that whenever a block reward is earned the pool distributes a fixed fraction $X_i$ of the reward to the $i$th pool share preceding the block share, where $i \geq 0$. 
\end{definition*}

\begin{definition*}[Allocation rule]
An \textbf{allocation rule} is a function $\mathcal{A}(t, h_t) = \{X^{(t, h_t)}_i\}_{0 \leq i \leq t}$ where $X^{(\sigma)}_i$ is a random variable denoting the value allocated to the contributor of the pool's $i$th share when the pool wins a block reward $B$ in state $\sigma = (t, h_t)$. 
\end{definition*}

\begin{definition*}[Pool share utility]
Given discount parameter $\delta$ and utility function $u$, the \textbf{discounted expected utility of a pool share} is:

$$U =  \sum_{i \geq 0} E[u(X_i)] \delta^i$$

\end{definition*}

\begin{definition*}[Steady-state fairness]
A pool strategy is \textbf{steady-state fair} if the sequence $\{U_k\}$ converges in $\mathbb{R}$, where $U_k$ denotes the expected utility of the $k$th pool share. The limit point of $\lim_{k \rightarrow \infty} U_k$ is the \textbf{steady-state utility} of the pool. 
\end{definition*}

\begin{definition*}[Steady-state optimality]
A pool strategy $\mathfrak{p}$ is \textbf{steady-state optimal} for a class $\mathcal{C}$ of steady-state fair pool strategies if and only if
$ \mathfrak{p} \in \argmax_{x \in \mathcal{C}} \lim_{k \rightarrow \infty} E[U(x)_k]$.
\end{definition*}

\begin{definition*}[Finite-window pool] 
A pool has \textbf{finite window n} if its allocation rule $\mathcal{A}(t, h_t) = \{X^{(t, h_t)}_i\}_{0 \leq i \leq t}$ satisfies $X^{(t, h_t)}_i = 0$ for all $i \leq t - n$. A \textbf{finite-window pool} has a finite window n for some $n \in \mathbb{N}$. 
\end{definition*} 

\begin{definition*}[$\epsilon$-approximation pool]
Let $\mathfrak{p}$ and $\mathfrak{p}'$ be two pools with $k$th share utility $U_k$ and $U'_k$ respectively. The pool $\mathfrak{p}'$ \textbf{$\epsilon$-approximates} the pool $\mathfrak{p}$ if $|U_k  - U'_k| \leq \epsilon$ for all $k \geq 1$. 
\end{definition*}	


\end{document}